\documentclass[%
 reprint,
 superscriptaddress,
showpacs,preprintnumbers,
 amsmath,amssymb,
 aps,prc,
floatfix,
]{revtex4-1}

\usepackage{amsmath} 
\usepackage{amssymb} 
\usepackage{array} 
\usepackage{bm}
\usepackage{color} 
\usepackage{comment} 
\usepackage{dcolumn}
\usepackage{epstopdf} 
\usepackage[T1]{fontenc} 
\usepackage{graphicx}
\usepackage{lmodern} 
\usepackage{longtable} 
\usepackage{mhchem} 
\usepackage{microtype} 
\usepackage{rotating} 
\usepackage{silence}
\usepackage{siunitx} 
\usepackage{subfigure} 
\usepackage{tabu} 
\usepackage{textcomp} 
\usepackage{threeparttablex} 
\usepackage{todonotes}
\usepackage{verbatim} 
\usepackage{wrapfig}
\usepackage[colorlinks=true]{hyperref}
\hypersetup{colorlinks=true, urlcolor=blue, citecolor=blue, linkcolor=blue, breaklinks=true}
\usepackage{physics}
\usepackage[normalem]{ulem}

\DisableLigatures[f]{encoding = *, family = * } 

\begin{document}

\title{Absence of Low-Energy Shape Coexistence in $^{80}$Ge: \\
The Nonobservation of a Proposed Excited 0$_2^+$ Level at 639~keV}

\author{F.H.~Garcia}
\affiliation{Department of Chemistry, Simon Fraser University, Burnaby, British Columbia V5A 1S6, Canada}

\author{C.~Andreoiu}
\affiliation{Department of Chemistry, Simon Fraser University, Burnaby, British Columbia V5A 1S6, Canada}

\author{G.C.~Ball}
\affiliation{TRIUMF, 4004 Wesbrook Mall, Vancouver, British Columbia V6T 2A3, Canada}

\author{A.~Bell}
\affiliation{Department of Chemistry, Simon Fraser University, Burnaby, British Columbia V5A 1S6, Canada}

\author{A.B.~Garnsworthy}
 \affiliation{TRIUMF, 4004 Wesbrook Mall, Vancouver, British Columbia, V6T 2A3, Canada}

\author{F. Nowacki}
\affiliation{Universit{\' e} de Strasbourg, IPHC, 23 rue du Loess 67037 Strasbourg, France}
\affiliation{CNRS, UMR7178, 67037 Strasbourg, France}

\author{C.M.~Petrache} 
\affiliation{ Universit{\' e} Paris-Saclay, CNRS/IN2P3, IJCLab, 91405 Orsay, France}

\author{A.~Poves}
\affiliation{Departamento de F\'{i}sica Te\'{o}rica and IFT$_{}$UAM/CSIC, Universidad Aut\'{o}noma de Madrid, 28049 Madrid, Spain}

\author{K.~Whitmore}
\affiliation{Department of Chemistry, Simon Fraser University, Burnaby, British Columbia V5A 1S6, Canada}

\author{F.A.~Ali}
\affiliation{Department of Physics, University of Guelph, Guelph, Ontario, N1G 2W1, Canada}
\affiliation{Department of Physics, College of Education, University of Sulaimani, P.O. Box 334, Sulaimani, Kurdistan Region, Iraq}


\author{N.~Bernier}
\altaffiliation{Present address: Department of Physics, University of the Western Cape, P/B X17, Bellville, ZA-7535 South Africa}
\affiliation{TRIUMF, 4004 Wesbrook Mall, Vancouver, British Columbia, V6T 2A3, Canada}
\affiliation{Department of Physics and Astronomy, University of British Columbia, Vancouver, British Columbia V6T 1Z4, Canada}

\author{S.S.~Bhattacharjee}
\affiliation{TRIUMF, 4004 Wesbrook Mall, Vancouver, British Columbia, V6T 2A3, Canada}

\author{M.~Bowry}
\altaffiliation{Present address: School of Computing, Engineering and Physical Sciences, University of the West of Scotland, Paisley PA1 2BE, United Kingdom}
\affiliation{TRIUMF, 4004 Wesbrook Mall, Vancouver, British Columbia, V6T 2A3, Canada}

\author{R.J.~Coleman}
\affiliation{Department of Physics, University of Guelph, Guelph, Ontario N1G 2W1, Canada}

\author{I.~Dillmann}
\affiliation{TRIUMF, 4004 Wesbrook Mall, Vancouver, BC, V6T 2A3, Canada}
\affiliation{Department of Physics and Astronomy, University of Victoria, Victoria, British Columbia V8P 5C2, Canada}

\author{I.~Djianto}
\affiliation{Department of Chemistry, Simon Fraser University, Burnaby, British Columbia V5A 1S6, Canada}

\author{A.M.~Forney}
\affiliation{Department of Chemistry and Biochemistry, University of Maryland College Park, College Park, Maryland 20742, USA}

\author{M.~Gascoine}
\affiliation{Department of Chemistry, Simon Fraser University, Burnaby, British Columbia V5A 1S6, Canada}

\author{G.~Hackman}
\affiliation{TRIUMF, 4004 Wesbrook Mall, Vancouver, British Columbia, V6T 2A3, Canada}

\author{K.G.~Leach}
\affiliation{Department of Physics, Colorado School of Mines, Golden, Colorado 80401, USA}

\author{A.N.~Murphy}
\affiliation{TRIUMF, 4004 Wesbrook Mall, Vancouver, British Columbia, V6T 2A3, Canada}

\author{C.R.~Natzke}
\affiliation{TRIUMF, 4004 Wesbrook Mall, Vancouver, British Columbia, V6T 2A3, Canada}
\affiliation{Department of Physics, Colorado School of Mines, Golden, Colorado 80401, USA}

\author{B.~Olaizola} 
\affiliation{TRIUMF, 4004 Wesbrook Mall, Vancouver, British Columbia, V6T 2A3, Canada}

\author{K.~Ortner} 
\affiliation{Department of Chemistry, Simon Fraser University, Burnaby, British Columbia V5A 1S6, Canada}

\author{E.E.~Peters} 
\affiliation{Department of Chemistry, University of Kentucky, Lexington, Kentucky 40506-0055, USA}

\author{M.M.~Rajabali} 
\affiliation{Department of Physics, Tennessee Technological University, Cookeville, Tennessee 38505, USA}

\author{K.~Raymond} 
\affiliation{Department of Chemistry, Simon Fraser University, Burnaby, British Columbia V5A 1S6, Canada}

\author{C.E.~Svensson}
\affiliation{Department of Physics, University of Guelph, Guelph, Ontario N1G 2W1, Canada}

\author{R.~Umashankar}
\affiliation{TRIUMF, 4004 Wesbrook Mall, Vancouver, British Columbia, V6T 2A3, Canada}

\author{J.~Williams} 
\altaffiliation{Present address: TRIUMF, 4004 Wesbrook Mall, Vancouver, British Columbia, V6T 2A3, Canada}
\affiliation{Department of Chemistry, Simon Fraser University, Burnaby, British Columbia V5A 1S6, Canada}

\author{D.~Yates} 
\affiliation{TRIUMF, 4004 Wesbrook Mall, Vancouver, British Columbia, V6T 2A3, Canada}
\affiliation{Department of Physics and Astronomy, University of British Columbia, Vancouver, BC V6T 1Z4, Canada}

\date{\today}

\begin{abstract}

The $^{80}$Ge structure was investigated in a high-statistics $\beta$-decay experiment of $^{80}$Ga using the GRIFFIN spectrometer at TRIUMF-ISAC through $\gamma$, $\beta$-$e$, $e$-$\gamma$ and $\gamma$-$\gamma$ spectroscopy. No evidence was found for the recently reported 0$_2^{+}$ 639-keV level suggested as evidence for low-energy shape coexistence in $^{80}$Ge. Large-scale shell model calculations performed in $^{78,80,82}$Ge place the $0^{+}_{2}$ level in $^{80}$Ge at 2\,MeV. The new experimental evidence combined with shell model predictions indicate that low-energy shape coexistence is not present in $^{80}$Ge.
\end{abstract}

\maketitle
Shape coexistence is ubiquitous across the chart of nuclides~\cite{Morinaga1956,Heyde2011,Garrett2016, Wood2012}, but found mainly in the vicinity of shell and subshell closures. It manifests as the appearance of two or more quantum states of different intrinsic shapes located within a narrow energy range. 
A key signature of shape coexistence in even-even nuclei is the presence of low-lying excited 0$^+$ states above the $0^{+}$ ground state. In most cases, these 0$^+$ states are connected by strong electric monopole transitions ($E0$), indicating significant mixing between different nuclear shapes with large differences in deformation. The microscopic origin of these 0$^+$ states is particle-hole excitations across a shell or subshell gap. The significant energy required to promote a pair of nucleons is offset by a large gain in correlation energy from the residual proton-neutron interaction~\cite{Wood2012}. 
Shape coexistence at the neutron-rich $Z = 28, N = 50$ doubly magic shell closure has been experimentally investigated in a spectroscopic study of $^{78}$Ni via in-beam $\gamma$-ray spectroscopy~\cite{Taniuchi2019}. The high-energy 2$_1^+$ excited state and a low-lying second 2$_2^+$ state separated by 0.31 MeV suggest that shape coexistence is present in $^{78}$Ni~\cite{Taniuchi2019}.

A number of state-of-the-art theoretical calculations using modern approaches have been performed for the $N = 50$ region including doubly magic $^{78}$Ni. These include $ab$ $initio$ approaches and the beyond-mean-field random-phase approximation~\cite{Taniuchi2019} as well as large-scale shell-model calculations~\cite{Nowacki2016} employing various phenomenological shell-model interactions. These calculations are in agreement with the experimental data now available for $^{78}$Ni, showing that the doubly magic nature is preserved, and a well-deformed prolate band is present at low excitation energy, representing a dramatic example of shape coexistence far from the valley of stability. Additionally, the phenomenological shell-model calculations predict a rapid transition from spherical ground states in the Ni isotopes up to $^{78}$Ni and deformed ground states for more neutron-rich isotopes~\cite{Taniuchi2019}.

The structural evolution of the Ge and Se nuclei from $N = 34-62$ has been studied within the interacting boson model (IBM)~\cite{Nomura2017}. In general, the IBM calculations agree with the trends in the experimental excitation energies for low-lying 0$^+$, 2$^+$, and 4$^+$ levels, including shape coexistence observed near $N = 40$, an increase in the excitation energies at $N = 50$, and also predict the onset of shape coexistence beyond $N = 52$. 

New experimental results for $^{80}$Ge, located two neutrons below the $N=50$ shell closure, were recently reported from a study performed at the ALTO facility using the $\beta$ decay of $^{80}$Ga to perform $\beta$-delayed electron-conversion spectroscopy~\cite{Gottardo2016}. A conversion electron peak at 628~keV was reported and attributed to the decay of a 0$^+_2$ state at 639~keV in $^{80}$Ge, located just below the first excited 2$^+$ state at 659~keV.

A comparison of the experimental energies of the 0$_2^+$ states in the $N=48$ isotones with phenomenological estimates from mass data was used to show lowering of the 0$_2^+$ states at $Z=32$ due to the pairing, monopole, and quadrupole terms of the interactions. Based on this analysis, the proposed 0$_2^+$ state in $^{80}$Ge was interpreted as a $\nu (2p-2h)$ excitation across the $N=50$ shell gap~\cite{Gottardo2016}; evidence of shape coexistence in $^{80}$Ge.

In the present work, confirmation of the existence of the $0_{2}^{+}$ state and shape coexistence in $^{80}$Ge was sought. States in $^{80}$Ge were studied via $\beta$ decay of $^{80}$Ga using conversion-electron and $\gamma$-ray spectroscopy. No experimental evidence for the previously proposed $0_2^+$ 639-keV level was found. Large-scale shell model calculations support this finding, and suggest that the $0_2^+$ level may be located near 2~MeV. These also agree with the theoretical trend found in the IBM~\cite{Nomura2017},but contradict the recent IBM-2 calculations~\cite{Zhang2018}.

The experiment was conducted at the Isotope Separator and ACcelerator (ISAC) facility~\cite{Dilling2014} at TRIUMF, where radioactive beams are produced via the Isotope Separation On-Line method. A 9.8 $\mu$A beam of protons was accelerated to 480~MeV by the main cyclotron and impinged onto a thick $\mathrm{UC}_{x}$ target, inducing spallation, fragmentation, and fission reactions. The Ga atoms of interest that did not diffuse from the production target were ionized using the ion-guide laser ion source~\cite{Raeder2014} which also strongly suppressed the surface ionized $^{80}$Rb isobaric contamination. An $A = 80 $ beam at 30~keV was selected by a high-resolution mass separator and sent to the experimental area. The resulting beam composition was $\sim$22$\%$ $^{80}$Ga and 78$\%$ $^{80}$Rb.

The $2\times10^{4}$~pps $^{80}\mathrm{Ga}$ beam was delivered to the Gamma-Ray Infrastructure For Fundamental Investigations of Nuclei (GRIFFIN)~\cite{Svensson2014,Rizwan2016,Garnsworthy2017,Garnsworthy2019} and implanted onto a Mylar tape system at the center of the GRIFFIN spectrometer. GRIFFIN is an array of up to 16 BGO compton-suppressed high-purity germanium (HPGe) clover detectors used for $\gamma$-ray detection and operated using a digital data acquisition system~\cite{Garnsworthy2017} in a triggerless mode. Only 15 HPGe clovers were used in the present work. GRIFFIN was operated in its optimal peak-to-total configuration~\cite{Garnsworthy2019} with the HPGe detectors located 14.5~cm from the beam implantation point, with an efficiency of 7\% at 1332 keV in clover addback mode.

Electrons produced by internal conversion  were detected using the Pentagonal Array of Conversion Electron Spectrometers (PACES)~\cite{Garnsworthy2019}. The array consists of five lithium-drifted silicon detectors, cooled with liquid nitrogen. The centers of the PACES detectors were located 3.15~cm from the implantation point, with an array efficiency $\sim2\%$. A single plastic scintillator, with an efficiency of $\sim$40\%, was positioned behind the implantation location at zero degrees to the beam axis for the tagging of $\beta$ particles~\cite{Garnsworthy2019}. A 10 mm thick Delrin absorber was placed around the vacuum chamber to prevent high-energy $\beta$ particles from reaching the surrounding HPGe detectors and limit bremsstrahlung~\cite{Garnsworthy2019}.

Tape cycles were chosen to maximize the implantation time and total decays of $^{80g,m}\mathrm{Ga}$ [$T_{1/2, gs}$=1.9(1)~s, $T_{1/2,m}$=1.3(2)~s] while reducing the activity from the subsequent decay of $^{80}\mathrm{Ge}$ ($T_{1/2}$=29.5~s) as well as the decay of the $^{80}\mathrm{Rb}$ contaminant ($T_{1/2}$=33.4~s). A typical cycle consisted of tape movement for 1.5~s, background measurement for 1.0~s, beam implantation for 15~s, and beam decay for 10~s. After each cycle, the implantation point on the tape was moved into a lead-shielded box outside of the spectrometer to reduce the background. Coincident hits from HPGe crystals within the same clover detector recorded within a 250~ns time window were combined into a single event to construct addback $\gamma$-ray events.

The efficiency of GRIFFIN was determined for the 81-keV to 3.2-MeV energy region using standard sources of $^{133}\mathrm{Ba}$, $^{152}\mathrm{Eu}$, $^{56}\mathrm{Co}$, and $^{60}\mathrm{Co}$. Summing corrections for $\gamma$-ray intensities were made by using a 180\textdegree\ $\gamma$-$\gamma$ coincidence matrix as described in Ref.~\cite{Garnsworthy2019}.

The $^{80}$Ge level scheme was constructed by setting gates on the time-random background-subtracted $\gamma$-$\gamma$ addback matrix. The comprehensive structure and spectroscopic information for $^{80}$Ge, including $\gamma$-ray intensities, branching ratios, angular correlations, $\beta$-decay lifetime and fast $\gamma$-ray lifetime measurements will be discussed in a forthcoming paper~\cite{Garcia2020}. All $\gamma$ rays and levels presented in Ref.~\cite{Verney2013} were observed in the current work, confirming the presence of both the 6$^{-}$ ground state and 3$^{-}$ isomer in \ce{^{80}Ga}. 

\begin{figure}
\centering
\includegraphics[width=\columnwidth]{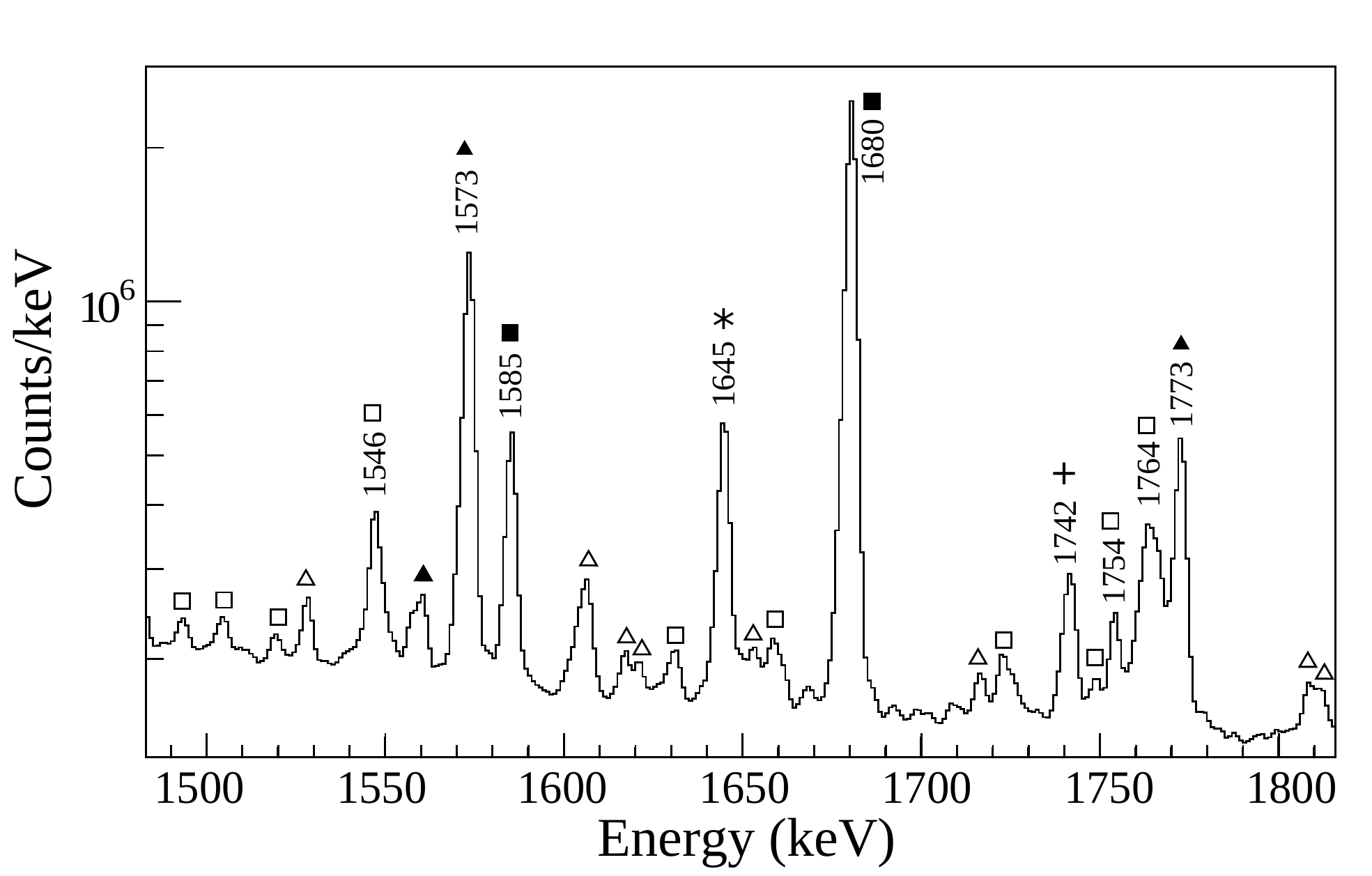}
\caption{\label{fig:gamma} The $\gamma$-ray spectrum in the energy range from 1500--1800 keV. The labels represent the following: the solid (open) squares are  known (new) transitions from levels fed by the 6$^{-}$ ground state of $^{80}$Ga; the solid (open) triangles are known (new) transitions from levels fed by the 3$^{-}$ isomer in $^{80}$Ga, the asterisk ($\ast$) is a transition in $^{80}$Se, the cross ($+$) is the sum peak between the strong 659.1- and 1083.4-keV transitions.}
\end{figure}

A portion of the $\gamma$-ray spectrum is shown in Fig.~\ref{fig:gamma}. Previously, Verney \textit{et al.} \cite{Verney2013}, using a beam of $^{80}$Ga produced by the photo-fission of UC$_{x}$, observed an increase in the relative intensity of $\gamma$ rays from low-lying states in $^{80}$Ge associated with the 6$^{-}$ ground state decay of $^{80}$Ga and a corresponding decrease in the relative intensity of the $\gamma$ rays associated with the 3$^{-}$ isomer decay, when compared with those obtained from a $^{80}$Ga beam produced by the thermal neutron fission of $^{235}$U studied by Hoff and Fogelberg~\cite{HoffFogelberg}. In the present Letter, similar but larger differences were observed, indicating that the 3$^{-}$ isomer content is different in all three cases. This difference can be used to estimate the 3$^{-}$ isomer content of the beam in the present work.  Specifically, the 2$^{+}$ 1573-keV state can only be directly fed by the 3$^{-}$ isomer, and the 8$^{+}$  3445 keV state is directly fed only by the 6$^{-}$ ground state. Comparing the $\beta$-feeding intensities of these two states, determined from relative $\gamma$-ray intensities, with the previous work results in an increase of 1.55(6) and a decrease of 0.66(3) for the 3445-keV and 1573-keV levels, respectively. This corresponds to a 3$^{-}$ isomeric content of 41(3)\% in the present work and 62(4)\% in the $^{80}$Ga beam produced by thermal neutron fission~\cite{HoffFogelberg} and ENSDF~\cite{NNDC}. From a comparison of $\beta$-feeding intensities for all levels observed in both experiments, calculated from the relative $\gamma$-ray intensities and assuming no $\beta$ feeding to the ground state of $^{80}$Ge, 13 levels were clearly identified as being fed by the 3$^{-}$ isomer; representing 46(2)\% of the total $\beta$-feeding intensity in the present work and 62(5)\% in Refs.~\cite{HoffFogelberg, NNDC} (see Ref.~\cite{Garcia2020}). Relative $\gamma$-ray intensities were not reported by Verney \textit{et al.}~\cite{Verney2013}, however, from the data shown in Fig. 2 of Ref.~\cite{Verney2013}, it is estimated that the $3^{-}$ isomer content in the $^{80}$Ga beam produced by photofission is $\sim 52$\%.

\begin{figure}
\centering
\includegraphics[width=\columnwidth]{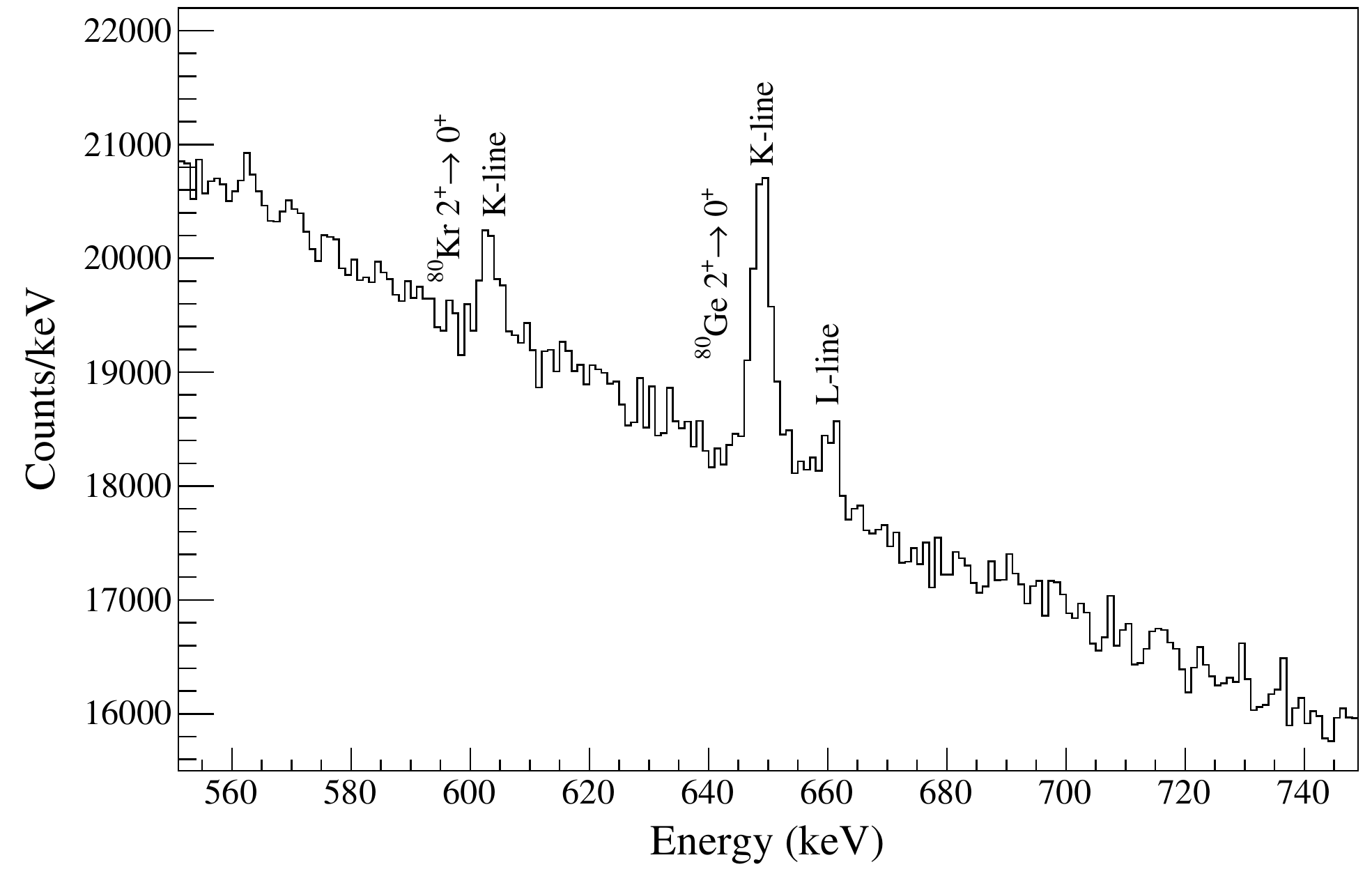}
\caption{\label{fig:paces} $\beta$-gated electron spectrum obtained following the $\beta$-decay of $^{80}$Ga showing the $2_1^+ \rightarrow 0_1^+$ \textit{K} line at 648~keV and \textit{L} line at 658~keV in $^{80}$Ge. The peak at 601~keV corresponds to the $2_1^+ \rightarrow 0_1^+$ \textit{K} line in $^{80}$Kr from the decay of $^{80}$Rb present in the beam. There is no evidence for the peak at 628 keV as reported by Gottardo \textit{et al.}~\cite{Gottardo2016}}.
\end{figure}

A portion of the $\beta$-gated electron spectrum is shown in Fig.~\ref{fig:paces}. The strong \textit{K} line from the $2^{+}_{1}\rightarrow0^{+}_{1}$ transition in $^{80}$Ge is clearly visible at 648~keV, along with the \textit{L} line at 658~keV. The \textit{K} line from the $2^{+}_{1}\rightarrow0^{+}_{1}$ decay in $^{80}\mathrm{Kr}$, populated by the $\beta$ decay of $^{80}\mathrm{Rb}$, is also visible at 601~keV. The ratio of the intensity of these two $K$ electron lines agrees with the value predicted from the ratio of the measured intensities of the corresponding $\gamma$-rays, corrected for internal conversion~\cite{Bricc}. No other significant features were observed in the rest of the spectrum outside the energy range shown in Fig.~\ref{fig:paces}. There is no evidence for a peak near 628~keV where the $0_2^+\rightarrow0_1^+$ $E0$ transition in $^{80}$Ge was previously reported~\cite{Gottardo2016}. 

Since 96\% of all $\beta$ decays from the 3$^{-}$ isomer and 6$^{-}$ ground state of $^{80}$Ga emit a 659-keV $\gamma$ ray, the intensity of the 648-keV $K$ electron line in Fig. 2 is proportional to the total $^{80}$Ga $\beta$ decays observed in the present experiment. Integrating the region centered around 628 keV and comparing the intensity of a hypothetical $E0$ transition to the intensity of the 648-keV $K$ line, corrected for the 41\% $3^{-}$ isomeric content of the beam and for internal conversion~\cite{Bricc}, yields a 2$\sigma$ limit of $<$ 0.02 per 100 decays of the 3$^{-}$ isomer in  $^{80}$Ga. By analogy, from the data presented in Fig. 2 of Ref. \cite{Gottardo2016} for a $^{80}$Ga $3^{-}$ isomeric component of 52\%, the intensity of the observed 628-keV electron peak is estimated to be $\sim$0.08(2) per 100 decays, a factor of four times the 2$\sigma$ upper limit and comparable to the intensity of the 601-keV $^{80}$Kr $K$ line observed in the present experiment. While there is no explanation for this discrepancy, it must be noted that the slope of the background in the $\beta$-gated electron spectrum shown in Fig. 2 of Gottardo \textit{et al.}~\cite{Gottardo2016} has an unusual shape that is not typical of Si(Li) detectors used in direct view of a $\beta$-decay source with a high $Q$ value emitting one or more $\gamma$ rays per decay~\cite{siegbahn1965}.

The 1764-keV $\gamma$ ray that was assigned by Gottardo \textit{et al.} \cite{Gottardo2016} to decay from a 2403-keV 2$^{+}$ state in  $^{80}$Ge to the proposed 639-keV $0_{2}^{+}$ level can be seen in the $\gamma$-ray spectrum in Fig.~\ref{fig:gamma}. The peak at 1742 keV has been identified as the sum peak of the intense 659- and 1083- keV transitions, and not from $^{80}$As as suggested in Ref.~\cite{Verney2013}. An unresolved peak at 1768~keV representing the summing of the 659- and 1109-keV transitions is also present in this spectrum, with an intensity of 40$\%$ of the 1742-keV peak.

\begin{figure}
\centering
\includegraphics[width=\columnwidth]{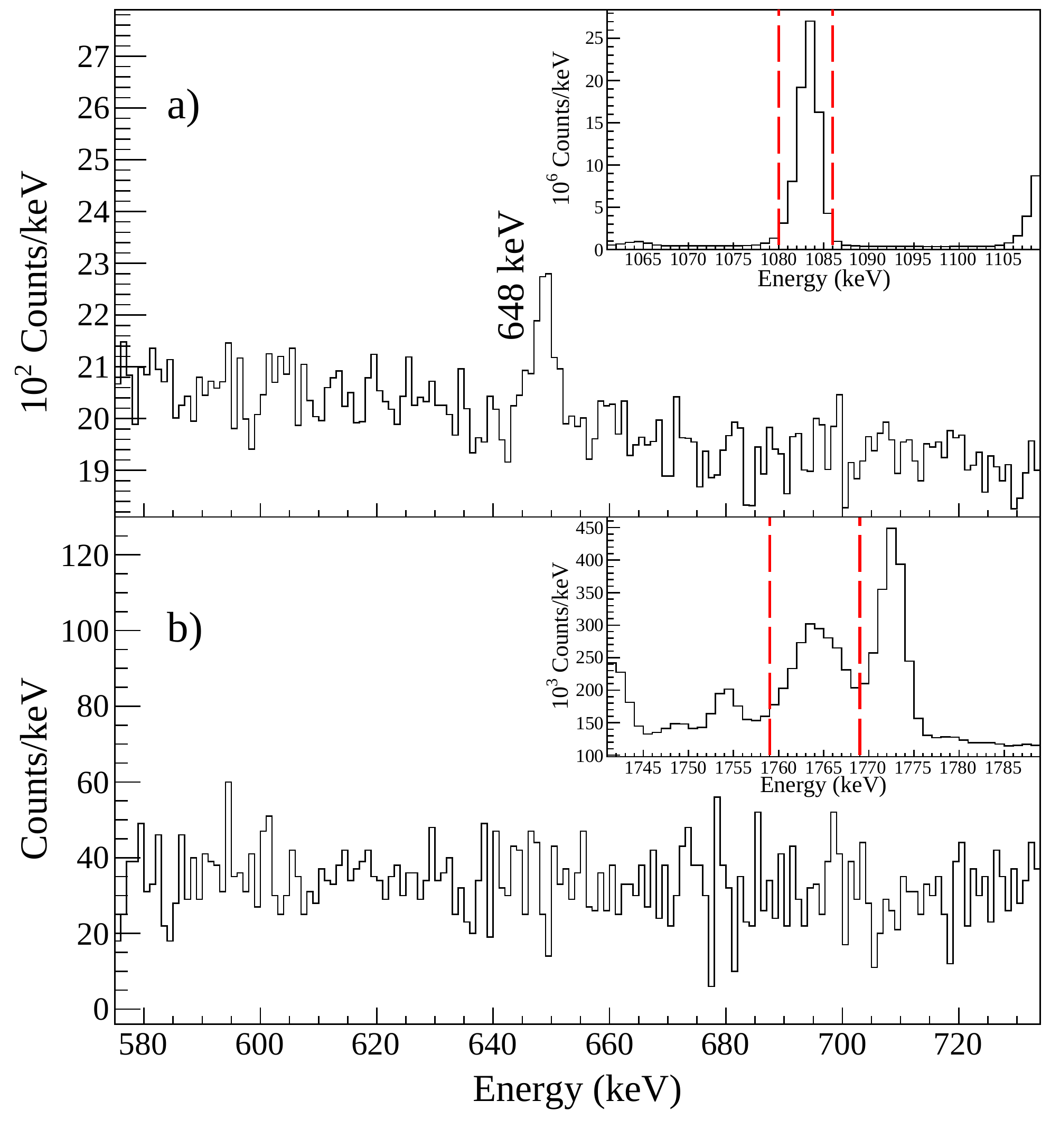}
\caption{\label{fig:1764paces}Background-corrected electron spectra, obtained from a $\gamma$-electron matrix by placing a) a gate on the 1083-keV transition, showing only the 648-keV $K$ line corresponding to the 659-keV transition, and b) a gate on the wide $\gamma$-ray peak in the 1764 keV region showing the absence of an electron line at 628~keV. The insets show the locations of the $\gamma$-ray gates.}
\end{figure}

Gates were placed on the $\gamma$-ray peaks in the $e-\gamma$ matrix to look for a 1764-keV $\gamma$-ray transition in coincidence with a 628-keV electron peak. A gate placed on the $4^{+}_{1} \rightarrow 2^{+}_{1}$ 1083-keV $\gamma$-ray generated the coincidence electron spectrum shown in Fig.~\ref{fig:1764paces}(a) which shows the \textit{K} line at 648 keV associated with the 659-keV $2^{+}_{1} \rightarrow 0^{+}_{1}$ transition in $^{80}$Ge. Comparing the converted intensity of the 648-keV electron line in the 1083-keV $\gamma$-gated electron spectrum, with the intensity of the $\gamma$-singles 1083-keV transition, yields an $e-\gamma$ coincidence efficiency of 1.6(2)$\%$. A wide gate on the region around 1764 keV produced the spectrum shown in Fig.~\ref{fig:1764paces}(b); no peak near 628 keV is present in this electron spectrum. The 2$\sigma$ limit for an $E0$ transition at 628 keV, determined from this spectrum, corresponds to $< 0.2\%$ of the intensity of this broad $\gamma$-ray peak and is equivalent to a 1764-keV transition intensity from the proposed level at 2403 keV in $^{80}$Ge~\cite{Gottardo2016} of $<$ 0.01 per 100 decays of the 3$^{-}$ isomer in $^{80}$Ga. Furthermore, the ratio of the 2$\sigma$ intensity limit for the 1764-keV transition to the intensity of the 1773-keV transition from the decay of the 3515-keV level in $^{80}$Ge, fed by the $3^{-}$ isomer \cite{Garcia2020, Verney2013} is $<$ 0.003, compared with the value of 0.3 reported by Gottardo \textit{et al.} \cite{Gottardo2016}.

\begin{figure}
\centering
\includegraphics[width=\columnwidth]{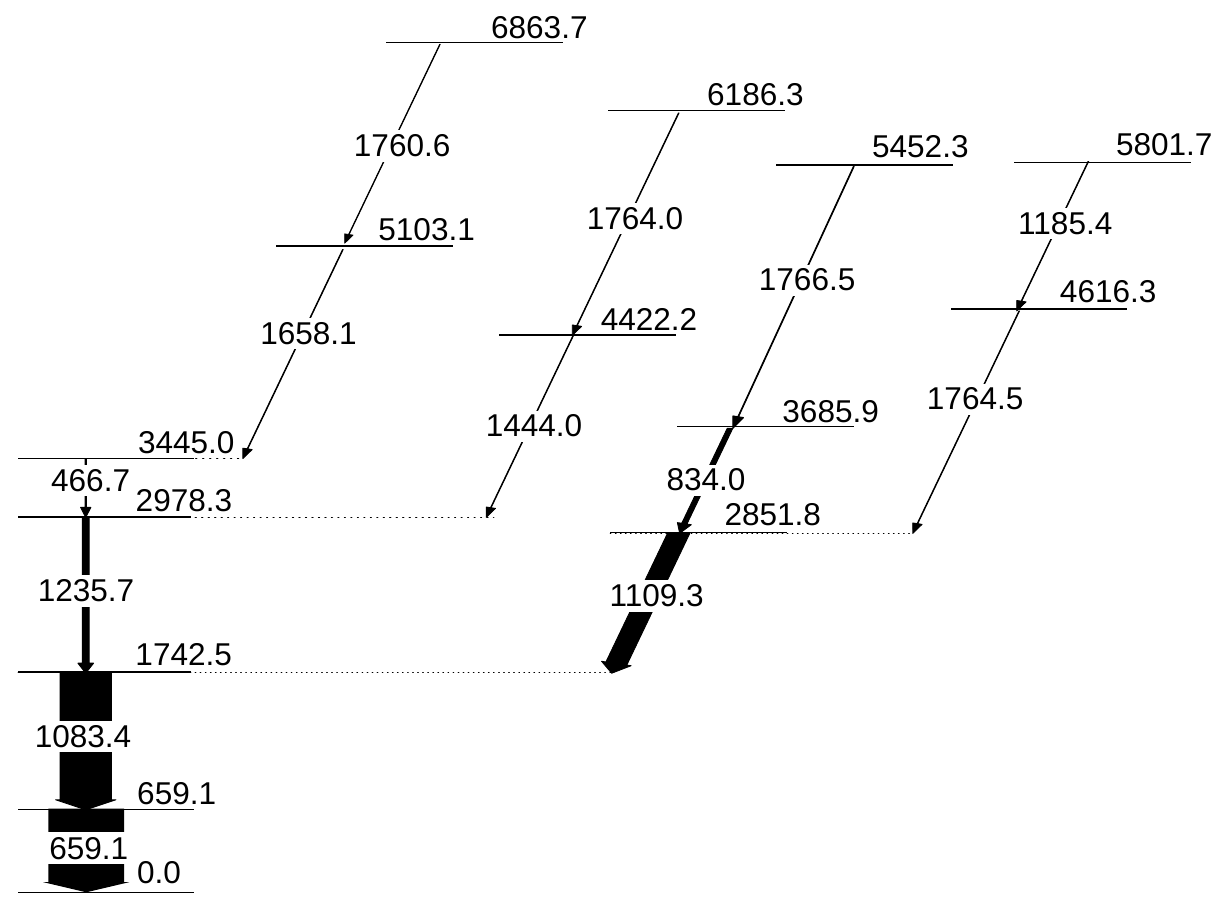}
\caption{\label{fig:LSP} Partial level scheme of $^{80}$Ge showing the placement of four new transitions that make up the wide peak around 1764~keV. The widths of the arrows are proportional to the relative intensities of the $\gamma$-ray transitions. }
\end{figure}

The unresolved $\gamma$ rays in the 1764-keV region seen in Fig.~\ref{fig:gamma} were further investigated by examining the $\gamma$-$\gamma$ coincidence relationships. By placing narrow gates in the 1764-keV region, distinct coincident spectra were observed, which were used to expand the $^{80}$Ge level scheme as shown in Fig.~\ref{fig:LSP}. Four new transitions were observed at 1760.6, 1764.0, 1764.5, and 1766.5 keV, all with intensities well below 1\%, relative to the 659-keV 2$^{+}_{1}\rightarrow 0^{+}_{1}$ transition. No evidence for the 2403-keV (2+) level or the 1764-keV transition from this state was found, as reported by Gottardo \textit{et al}.~\cite{Gottardo2016}.

Large-scale shell-model calculations with configuration interactions have been carried out to explore the nuclear structure around $N=50$ above $^{78}$Ni. Two valence spaces were considered. The first valence space, LNPS, is based on a $^{48}$Ca core and encompasses the full $pf$ shell for the protons and the $pf$-shell orbits above the 0$f_{7/2}$ plus the 0$g_{9/2}$ and 1$d_{5/2}$ orbitals for the neutrons. The effective interaction is the current version of the original LNPS \cite{Lenzi2010} which incorporates some minor changes which do not affect the predictions near $N=40$ and improved the behavior towards $N=50$. The second is the PF-SDG space, based on a $^{60}$Ca core and consisting of the $p=3$ major oscillator shell ($pf$) for the protons and the $p=4$ major oscillator shell ($sdg$) for the neutrons. The PF-SDG interaction used in this work is the one described in Ref.~\cite{Nowacki2016}. In addition to \ce{^{80}Ge}, calculations have been performed for the neighboring isotopes \ce{^{78}Ge} and \ce{^{82}Ge}, where excited $0^{+}_2$ states have been observed.

For the specific case of $^{82}$Ge, both interactions predict a 0$^{+}$ intruder state near 2 MeV, but the deformation extracted from the restricted valance space is small. This prediction agrees with the observed $0_{2}^{+}$ state at 2334 keV in $^{82}$Ge that has been attributed to a deformed rotational band in $^{82}$Ge resulting from $2p-2h$ excitations across the $N=50$ closed shell \cite{Hwang2011}. All of the other $0^{+}$ states predicted by both interactions for $^{78,80,82}$Ge arise from the recoupling of different valence particles. Additional intruder states likely exist at higher excitation energies but tracking them is computationally demanding and beyond the scope of this study.
\begin{figure}
\centering
\includegraphics[width=\columnwidth]{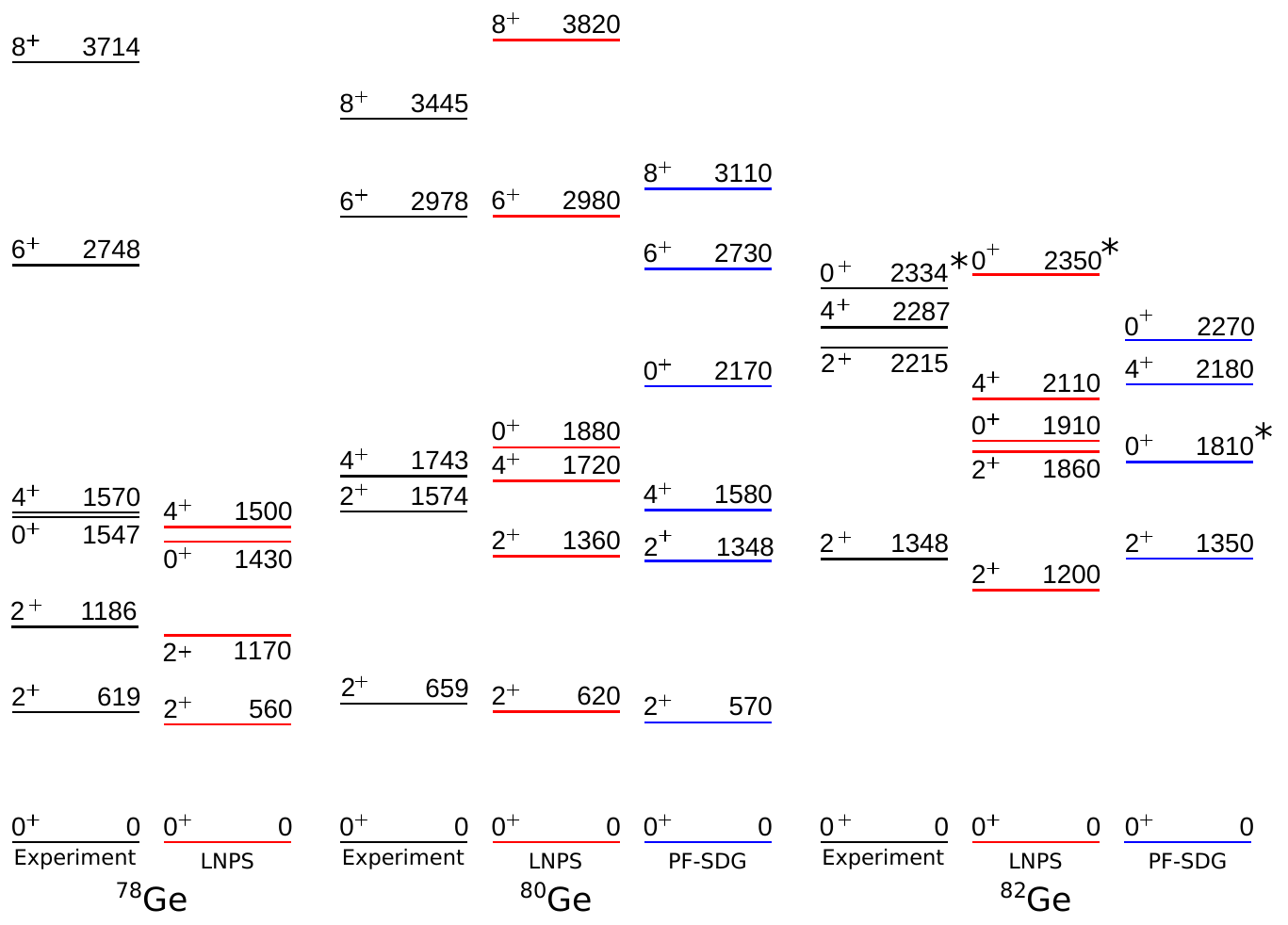}
\caption{\label{fig:evt} Comparison of experimental (black) and calculated nuclear levels in $^{78,80,82}$Ge using the LNPS (red) and PF-SDG (blue) interactions. The asterisks (*) identify intruderlike structures.}
\end{figure}

The results of these calculations can be seen in Fig.~\ref{fig:evt}. For each of the $^{78,80,82}$Ge isotopes, the energies of the low-lying positive-parity states are shown along with the calculated values. In all cases, the energies of the $2^{+}_{1}$ and $4^{+}_{1}$ levels are well reproduced. The calculation with the LNPS valence space also reproduces the $2^{+}_{2}$ levels. In both \ce{^{78}Ge} and \ce{^{82}Ge}, the $0^{+}_{2}$ levels are known to exist above 1.5~MeV, and the calculations predict these energies very well. In the case of \ce{^{80}Ge}, the calculations predict the $0_{2}^{+}$  state to be at a relatively high excitation energy near 2~MeV. Whether the $0_{2}^{+}$ state is observed near 2~MeV in the present work must wait for a more complete analysis of all the very weak $\gamma$-ray transitions observed in this high-statistics dataset~\cite{Garcia2020}.

In conclusion, the $\beta$ decay of $^{80}$Ga to $^{80}$Ge has been studied using the GRIFFIN spectrometer at TRIUMF-ISAC. The $^{80}$Ge nucleus has been investigated via $\gamma$-ray and conversion-electron spectroscopy. No evidence for an excited 0$_{2}^{+}$ state located below the 2$_{1}^{+}$ state at 659~keV is found in this experiment, despite detailed investigations using multiple $\beta$-electron, $\gamma$-electron and $\gamma$-$\gamma$ coincidences. Additionally, driven by these experimental results, large-scale shell-model calculations that reproduced well the excited $0^{+}_2$ states in $^{78,82}$Ge and other low-lying levels in $^{78-82}$Ge, cannot replicate the 0$_{2}^{+}$ state suggested at 639 keV in $^{80}$Ge; the calculations instead predict the first excited $0^{+}$ state at 2~MeV. We conclude that the $0^{+}_{2}$ level at 639-keV excitation energy reported by Gottardo \textit{et al.}~\cite{Gottardo2016} does not, in fact, exist in $^{80}$Ge and that this isotope does not exhibit low-energy shape coexistence.\\

\begin{acknowledgments}
\vspace{-10pt}
We would like to thank the operations and beam delivery staff at TRIUMF for providing the $^{80}$Ge radioactive beam. This work was supported in part by the Natural Sciences and Engineering Research Council of Canada. The GRIFFIN infrastructure has been funded jointly by the Canada Foundation for Innovation, the University of Guelph, TRIUMF, the British Columbia Knowledge Development Fund, and the Ontario Ministry of Research and Innovation. TRIUMF receives federal funding via a contribution agreement through the National Research Council Canada (NRC). This work was funded in part by the U.S. Department of Energy, Office of Science under Grant No. DE-SC0017649. This material is based upon work supported in part by the U.S. National Science Foundation under Grant No. PHY-1913028. C.E.S. acknowledges support from the Canada Research Chairs program. A.P. acknowledges the support of the Ministerio de Ciencia, Inovaci\'on y Universidades (Spain), Severo Ochoa Programme SEV-2016-0597 and Grant No. PGC-2018-94583. C.A. acknowledges the fruitful discussions with D. Verney on the spectroscopy of Ge isotopes.
\end{acknowledgments}
\bibliographystyle{apsrev}
\bibliography{LT17457-80Ge-FHG-PRLFinal}

\end{document}